\documentclass[a4paper,twoside]{article}

\usepackage{epsfig}
\usepackage{subcaption}
\usepackage{calc}
\usepackage{amssymb}
\usepackage{amstext}
\usepackage{amsmath}
\usepackage{amsthm}
\usepackage{multicol}
\usepackage{pslatex} 
\usepackage[authoryear]{natbib}
\usepackage[bottom]{footmisc}
\usepackage{url}
\usepackage[table]{xcolor}

\usepackage{SCITEPRESS}     % Please add other packages that you may need BEFORE the SCITEPRESS.sty package

\begin{document} 
\title{Web-based Search: How Do Animated User Interface Elements Affect Autistic and Non-Autistic Users?}

% \orcidAuthor{0000-0000-0000-0000} 
\author{\authorname{Alexandra L. Uitdenbogerd, Maria Spichkova, Mona Alzahrani }
\affiliation{School of Computing Technologies, RMIT University, Melbourne, Australia} 
\email{\{sandra.uitdenbogerd,maria.spichkova\}@rmit.edu.au, mona.s.alzhrani@gmail.com}
}
%s 
 
\keywords{Search, Information retrieval, Autistic users, Animation, Usability, HCI}

\abstract{
  Many websites and other user interfaces include animated elements, particularly for advertisements. However, these can have a negative impact on users,
  with some cohorts, such as autistic users, 
  being more affected.
  In our mixed methods study on the effect of irrelevant animations on usability we observed the effect on search 
  activities.
  For those greatly impacted by on-screen animation the effect was not always to slow down a task, but search terms were entered hastily to avoid more exposure, with shorter queries on average and a greater tendency to copy and paste during query formulation. 
  Autistic users found the task more mentally demanding, and were more distracted or annoyed by the animations.
\\
\emph{Preprint. Accepted to the 17th International Conference on Evaluation of Novel Approaches to Software
Engineering (ENASE 2022). Final version published by SCITEPRESS, http://www.scitepress.org}
}

\onecolumn \maketitle \normalsize \setcounter{footnote}{0} \vfill

%========================================
\section{\uppercase{Introduction}}

Animations form part of many user interfaces, particularly on websites. However, they are known to interfere with a user's ability to complete their tasks~\citep{hong2007web}.
There has been much research on the impact of animated banner advertisements on the user experience, with particular attention to potential sales~\citep[see for example][]{lohtia2003impact,rau2006study,yoo2004assessing,zorn2012impact}. There are fewer studies that examine how well users achieve their goals on web-sites in the presence of on-screen animations.
There is some evidence that users' ability to complete tasks is more impacted for browsing than search, with the former considered a cognitively more demanding task \citep{cheung2017effects,pagendarm2001users,wang2014eye,hong2007web}. %,

Some users may be more impacted than others by having animated elements in the user interface. Indeed, even for static pages, there is evidence that autistic users have more difficulty focusing on the relevant screen elements in order to achieve their goals~\citep{eraslan2019web}. 

According to~\citet{vos2016global}, approximately 1\% of the overall population might be on on the autism spectrum.  
Individuals diagnosed with an autism spectrum condition often have hypersensitivity or hyposensitivity to sensory inputs, in addition to atypical attention tendencies~\citep{schaaf2005occupational}. There is some evidence from eye-tracking experiments that autistic users are more likely to be distracted by irrelevant elements on a user interface~\citep{eraslan2019web}. However, despite moving content being stated as a barrier to accessibility\footnote{\url{https://www.w3.org/WAI/people-use-web/abilities-barriers/}}, current web development guidelines addressing accessibility for those with cognitive differences appear to be largely based on anecdotal evidence~\citep{seeman2015cognitive}.

\emph{Contributions:} In this paper we report on part of a study that examined the impact  of irrelevant on-screen animation on two cohorts of users' ability to complete a range of tasks \citep{MonaKES21,MonaICSE2022}. 
Here we report the results and additional analysis related to the search task from the study to answer the following research question:

\emph{\textbf{RQ:} How does the effect of irrelevant animations within user interfaces vary between autistic and non-autistic users,  while conducting Web search tasks?} 
 
 The aim of our work was to investigate the interaction between distractors and users’ cognitive style (autistic versus non-autistic), in addition to user performance across  various internet-based tasks. Together with wring and reading emails, search is deemed the most popular online activity~\citep{purcell2011search} and was therefore an essential task to include. % in our study.  

% \emph{Outline:} 
% The rest of the paper is organised as follows. 
% Section~\ref{sec:related} introduces the related work. 
% In Section~\ref{sec:experiments}, we discuss the methodology we applied to conduct the study. 
% In Section~\ref{sec:results}, we present the results of the study, which are then analysed and discussed in Section~\ref{sec:discussion}. The limitations of our study are discussed in Section~\ref{sec:limit}. Finally, Section~\ref{sec:concl} summarises the paper.

%========================================
\section{\uppercase{Related work}}
\label{sec:related}

In our preliminary work, we presented and discussed a literature survey on the animation case studies~\citep{MonaKES21}. 
A number of studies in various disciplines have focused on the influences of animations and images on non-autistic users, such as the work of~\citet{hong2007web}, whereas there are only very few studies on autistic users. To our best knowledge, none of the existing studies has focused on the analysis of the impact of animated user interface elements on conducting Web-based search tasks by autistic and non-autistic users.

The guidelines on Web accessibility have been provided by the World Wide Web Consortium (W3C) in the form of the Web Content Accessibility Guidelines (WCAG) Version 2.1. However, users with cognitive impairments confront challenges that were not focused on so much in WCAG, as catering to physically disabled users’ needs  was the main emphasis \citep{harper2019web}. 
 
According to the Weak Central Coherence theory \citep{happe2006weak}, the cognitive profile of autistic individuals tends to be biased toward local sensory information processing, rather than semantic, contextual and global information processing. Despite the local bias,~\citet{eraslan2019web} demonstrated that autistic users had more holistic (and therefore less focused) eye-tracking patterns than non-autistic users, leading to potentially lower success in some focused tasks.
Perhaps this is due to the high overlap between autism and Attention Deficit/Hyperactivity Disorder (ADHD) diagnoses~\citep{schaaf2005occupational}.
 
A review of user experience studies with autistic users was presented in \citet{ccorlu2017involving}. It covers 98 studies  conducted between 2010 and 2016. 
The authors applied both qualitative and quantitative approaches in their meta-analysis. Their results showed that the most studied cohort are children (47\%) with only 11.9\% of studies involving autistic adults. Most studies were focused on software to address issues such as social interaction and communication. Very few, if any, addressed accessibility issues of the Web or other general software.
 
A systematic literature review on the impact of technology on autistic individuals was presented in \citet{valencia2019impact}. The authors reviewed 94 studies to analyse 
how the use of technology in educational contexts helps autistic people develop several skills, how these approaches consider aspects of user experience, usability and accessibility, and how game elements are used to enrich learning environments. As with ~\citet{ccorlu2017involving}, the articles in this review were largely about educating autistic children, with accessibility and usability not adequately addressed.
Similarly, there are many other studies focused on technology solutions for autistic children \citep{battocchi2010collaborative,millen2010development,sitdhisanguan2012using,gentry2010personal}.
 
Results from an anonymous on-line survey on the user experience of software or technology designed for autistic people were discussed in \citet{putnam2008software} but the majority of respondents were parents or carers of autistic children rather than autistic users themselves. The emphasis in the survey results was therefore more on solving social, communication and educational problems rather than specific user interface issues.

In summary, there is evidence that irrelevant animation impacts users and that autistic users are more likely to be distracted by static irrelevant elements on a user interface. However, apart from our study, there does not appear to be other research into how irrelevant animation impacts autistic users during search tasks.
 
 %==============

\begin{figure*}
\begin{center}
  \includegraphics[width=0.9\textwidth]{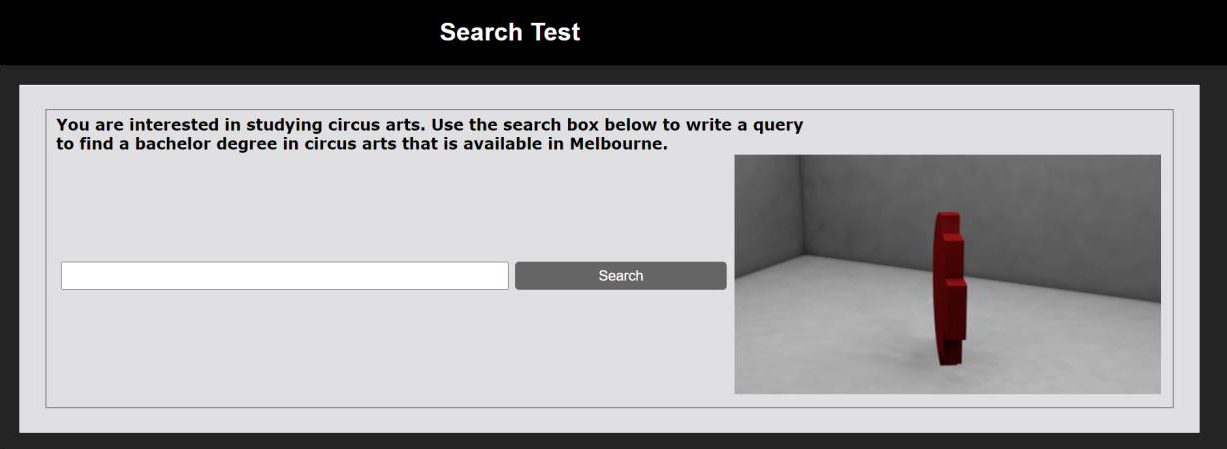}
 \end{center}
 \caption{Search Screenshot}
  %\Description{Search task and search textbox with an image distractor to the right of the screen.}
  \label{fig:searchScreen}
\end{figure*}

\section{\uppercase{Experiments}}
\label{sec:experiments}
In order to obtain comprehensive insights into users' experience, our approach was to use mixed methods, with quantitative and qualitative data collected in parallel. We triangulated based on the different types of evidence collected in the study.

%\subsection{Procedure}
Before commencing the study, approval was obtained from the relevant university ethics committee. Participants who answered recruitment advertisements and consented to take part initially completed a pre-questionnaire about their experiences and any diagnoses that may impact their participation, such as Attention Deficit/Hyperactivity Disorder (ADHD) or vision impairment.  

Our aim was to conduct the experiment with participants representing both autistic and non-autistic populations of users, where our inclusion criteria for classing a participant as autistic were a diagnosis of autism spectrum condition (ASC) levels 1 and 2 without intellectual disability, or self-identification as autistic. 

Three severity tiers of ASD (ASD level 3, ASD level 2 and ASD level 1) are defined in the most recent version of The Diagnostic and Statistical Manual of Mental Disorders (DSM-5)~\citep{american2013diagnostic}. As per DSM-5, average to superior intelligence is seen among individuals at ASD level 1 \emph{Autism Spectrum Condition} (ASC), although sensory sensitivity and social contexts may nevertheless be challenging for them.

 An on-line meeting was then held between the third author and the participant, during which the participant completed a set of six tasks, each with a different distractor. 
Given that users may become tired as the experiment progresses, it may be this factor that causes the final tasks to potentially be completed to a worse degree rather than the animation’s impact. Consequently, the results or findings of prior conditions may bias the user’s answers or reactions to the later conditions. In the case of animated elements, the learning effect means that participants learn to ignore these distractors.

\citet{hong2007web} state that when users initially see animation on a website they may find it difficult to ignore, but they will learn to partially block it and focus on the main task. To avoid this threat to validity, we organised our experiment using a Latin Square arrangement~\citep{fisher1938statistical}, which is a common approach for experimental  design~\citep{majrashi2016cross,burke2005high}. The Latin Square design is a technique for ordering conditions and tasks in a balanced way across study participants. In this case, the animation conditions were rotated across tasks, so that each user completed each task with a different animation condition and subsequent users completed them with different task-animation-condition combinations to prior users. 

Thus, each of the animation conditions consisted of a rotating  university logo, the size and speed of which were varied. We report here the results pertaining to the search task, the screen shot for which is shown in Figure~\ref{fig:searchScreen}. The order of animation is reversed after every six participants, as shown in Table \ref{tab:Latin}.

\begin{table}[ht!]
\centering
\caption{Latin Square design (\footnotesize{Case 1: Black and White image, Case 2: Slow rotate, Case 3: Coloured image, Case 4: Fast rotate, Case 5: Nothing and Case 6: Small slow rotate})} 
\label{tab:Latin}
\begin{tiny}
\begin{tabular}{|l|l|l|l|l|l|l|}
\hline
\textbf{} & \textbf{Task 1} & \textbf{Task 2} & \textbf{Task 3} & \textbf{Task 4} & \textbf{Task 5} & \textbf{Task 6} \\ \hline
\textbf{case 1 }                              
& User 1          & User 2          & User 3          & User 4          & User 5          & User 6          \\ \hline
\textbf{case 2}                                  & User 2          & User 3          & User 4          & User 5          & User 6          & User 1          \\ \hline
\textbf{case 3 }                                 & User 3          & User 4          & User 5          & User 6          & User 1          & User 2          \\ \hline
\textbf{case 4 }                                 & User 4          & User 5          & User 6          & User 1          & User 2          & User 3          \\ \hline
\textbf{case 5}                                  & User 5          & User 6          & User 1          & User 2          & User 3          & User 4          \\ \hline
\textbf{case 6 }                                 & User 6          & User 1          & User 2          & User 3          & User 4          & User 5          \\ \hline
\textbf{case 6}                                  & User 7          & User 8          & User 9          & User 10         & User 11         & User 12         \\ \hline
\textbf{case 5}                                  & User 8          & User 9          & User 10         & User 11         & User 12         & User 7          \\ \hline
\textbf{case 4}                                  & User 9          & User 10         & User 11         & User 12         & User 7          & User 8          \\ \hline
\textbf{case 3 }                                 & User 10         & User 11         & User 12         & User 7          & User 8          & User 9          \\ \hline
\textbf{case 2}                                  & User 11         & User 12         & User 7          & User 8          & User 9          & User 10         \\ \hline
\textbf{case 1 }                                 & User 12         & User 7          & User 8          & User 9          & User 10         & User 11    
 \\
\hline
\end{tabular}
\end{tiny}
\end{table}

~\\
The experiment was conducted online, with users’ screens being observed via online tools, such as screen sharing via Skype. To measure the animation’s effect on autistic and non-autistic users, the task completion rate, user error rate and time taken per task were recorded. User error rate was defined differently for each task. For search it was defined as a query with irrelevant words or a blank query. For the analysis in this paper, we use measures of query success, described in Section~\ref{dataCollection}.

After each task, the participant answered several questions related to the task they had just completed. A further questionnaire was answered after all tasks had been completed. In addition to the data collected via the web application and on-line survey forms, the on-line meeting was screen-recorded for later analysis.

%\subsection{Data Collection and Analysis}
%\label{dataCollection}

We posted an announcement on social media, with online autism organisations, as well as websites that aim to exchange surveys, to recruit participants for both the autistic and non-autistic cohorts.
Twelve autistic and twenty six non-autistic participants completed the study. 
All participants classed as autistic in this study stated that they had a diagnosis of autism. 

Task durations were captured, as well as participant responses regarding the perceived effort in completing the task. For the search task, the search query was recorded. All queries were later tested on a search engine to determine whether they were successful by recording the rank of the first relevant document. 

Two simple metrics were used to compare the different search conditions: Success at 10 and Mean Reciprocal Rank (MRR). The Success at 10 metric gives a score of 1 when there is a relevant query in the first 10 results and 0 otherwise. Thus, the mean across queries will be a number between 0 and 1. The metric was selected to represent whether the user's query was successful at all, given that a typical search engine provides ten results per page and most users will only look at the first results page most of the time~\citep{zhang2006some}.

MRR is a simple measure that uses the rank of the first relevant query result to represent relative success in the search task. For example, if the first relevant result was in position number 2, the reciprocal rank is 0.5. As results appear later in the ranked list, the reciprocal rank approaches zero.
The mean for both measures was calculated across participants' queries for the one search task rather than across multiple search tasks. 

We chose the DuckDuckGo search engine for query testing to minimise any personalisation that may occur during the search session. Location information was disabled in the browser during search.
Documents were classed as relevant if they referred to a bachelor degree on circus arts at a university located in Melbourne, as per the task presented in Figure~\ref{fig:searchScreen}.

In addition, query lengths are reported for each condition and cohort, as well as whether a query was typed or cut and pasted from the task description.

%======================================================
\section{\uppercase{Results}}
\label{sec:results}

\begin{table}[!htb]\centering
\footnotesize
    \caption{Task duration in seconds for both conditions (with and without animation)}
    \label{tab:durations}
    \begin{tabular}{llrrr
%        S[table-format=3.2]
%        S[table-format=3.2]
%        S[table-format=3.0]
        }
         \hline
        Type   of Task & Animation &
%        \multicolumn{1}{C{0.09\linewidth}}
        {$M$} &
%        \multicolumn{1}{C{0.09\linewidth}}
        {$SD$}   &
%        \multicolumn{1}{C{0.09\linewidth}}
        {$N$}    \\
         \hline
%        \multirow{3}{*}
        {Menu Options} & No   animation & 96.52  & 33.08  & 19  \\
                                          & Animation      & 109.36 & 44.99  & 19  \\
                                          & Total          & 102.94 & 39.49  & 38  \\
                                          &&&&\\
%        \multirow{3}{*}
        {Selection}        & No   animation & 74.10  & 30.11  & 19  \\
                                          & Animation      & 115.44 & 51.41  & 18  \\
                                          & Total          & 94.21  & 46.26  & 37  \\
                                          &&&&\\
%        \multirow{3}{*}
        {Search Query }            & No   animation & 59.73  & 26.18  & 19  \\
                  Formulation                        & Animation      & 55.68  & 22.30  & 19  \\
                                          & Total          & 57.71  & 24.07  & 38  \\
                                          %&&&&\\
         \hline
    \end{tabular}
\end{table}

\begin{table}[!htb]\centering
    \caption{Mean reciprocal rank}
    \label{tab:mrr}
    \footnotesize
    \begin{tabular}{lrr}
         \hline
& Animation & No Animation\\
         \hline
Autistic & 0.86 & 0.70\\
Non-Autistic & 0.88 & 0.94\\
         \hline
    \end{tabular}
\end{table}

%\vspace{5mm}

\begin{table}[!htb]\centering
    \caption{Mean query lengths in characters}
    \label{tab:queryLengths}
    \footnotesize
    \begin{tabular}{lrr}
        \hline
& Animation & No Animation\\
        \hline
Autistic & 30.1 & 31.0\\
Non-Autistic & 29.8 & 36.8\\
        \hline
    \end{tabular}
\end{table}

Table~\ref{tab:durations} shows the task duration across all participants for the search query formulation task and two browsing tasks. We use here the following notation: $N$ refers to the number of participants, $M$ the mean and $SD$ the standard deviation. Unlike the other tasks, search query formulation under the animation condition was slightly faster on average. Despite the speed, all queries completed in the animated condition were successful in retrieving a relevant result within the first ten documents with the DuckDuckGo search engine (as judged by the third author). The only unsuccessful query by this measure (``performance arts'') was formulated by an autistic user in the unanimated condition.

Table~\ref{tab:mrr} shows that the mean reciprocal rank only differed greatly for the autistic unanimated case, which was the impact of the single unsuccessful query mentioned above. Excluding the outlier changes the MRR from 0.7 to 0.88. It must be noted that this is a summary of only four values when the outlier is excluded. The vast majority of queries (30) had their first relevant result at rank 1, 6 had a rank of 2 and one a rank of 4. Interestingly, five of the six queries with rank 2 were in the animated condition, which may be why there is a notable difference between the animation and non-animation scores for the non-autistic cohort. The queries with rank 2 either left out the word Melbourne (for example, ``circus arts degre'') or used ``art'' instead of ``arts'' (for example, ``Melbourne circus performance art schools'').

Table~\ref{tab:queryLengths} shows the mean query lengths in characters for each of the cohorts and animation conditions. These means exclude one outlier, which was a pasted query of length 61 (`bachelor degree in circus arts that is available in Melbourne') formulated by a non-autistic user with the fast-rotating animation condition. With the outlier included, the mean is 32.2. The trend shows that in general, queries were longer in the unanimated condition, with the difference in query length being greater for non-autistic users.

Table~\ref{tab:pastedQueries} shows the proportion of queries that were pasted for each cohort and condition. 
More of those occurring in the animated condition were pasted. 
Pasted queries tended to be longer (mean query length was 37.9 characters, excluding the outlier) than for unpasted (mean query length was 31.2 characters).\\
~

\begin{table}[!htb]\centering
    \caption{Proportion of pasted queries}
    \label{tab:pastedQueries}
    \footnotesize
    \begin{tabular}{lrrr}
        \hline
& Animation & No Animation & All\\
        \hline
Autistic & 0.14 & 0.00 & 0.08\\
Non-Autistic & 0.31 & 0.23 & 0.27\\
Combined & 0.25 & 0.17 & 0.21\\
        \hline
    \end{tabular}
\end{table}

\subsection{Perceived Effort}
After the task, participants answered a four-question perceived effort questionnaire. Responses were on a scale from 1 to 100 and varied greatly between participants. These are summarised in Table~\ref{tab:effort}. 

Of note was that autistic participants under the animation condition averaged 40.3 for the question \emph{``How mentally demanding was the task''}, whereas all other cases had averages in the range 16.1 to 23.1. For the question \emph{``How successful were you in accomplishing what you were asked to do?''} autistic participant responses averaged 76.3 for the animation case and 71.2 in the unanimated case, whereas non-autistic participants averaged 92.8 and 99.4 respectively. 

For the question \emph{``How hard did you have to work to accomplish your level of performance?''}, non-autistic participants in the unanimated case averaged 4.5 in their responses, whereas all other cases had averages in the range 18.2 to 18.6. For the question \emph{``How insecure discouraged irritated stressed and annoyed were you?''}, both cohorts had similar scores in the unanimated case (11.2 for autistic participants and 10 for non-autistic), but autistic participants scored nearly double (37.9) that of non-autistic ones (20.9) when animation was present.

\begin{table}[!htb]\centering
    \caption{Perceived effort}
    \label{tab:effort}
    \footnotesize
    \begin{tabular}{p{2.5cm}lllr}
         \hline
        Question & Animation & Autistic & {Mean} \\
%        \multicolumn{1}{C{0.09\linewidth}}
         \hline
%        \multirow{3}{*}
        {How mentally} & 
        No  & No  & 16.1  \\
       demanding was  &   No    &  Yes  & 21.8  \\
%        \\
         the task?   & Yes      & No  & 23.1  \\
            &  Yes     &  Yes  & 40.3  \\
%        \multirow{3}{*}
 \hline
        {How successful }        & No  & No  & 99.4  \\
         were you in   &  No     &  Yes  & 71.2  \\
%            \\
         accomplishing     & Yes      & No  & 92.8  \\
          what you were  & Yes      &  Yes  & 76.3  \\
            asked to do? &        &    &    \\
 \hline
         How hard did you     & No  & No  & 4.5  \\
      have to work to      &  No     &  Yes  & 18.2  \\
       accomplish  your      & Yes      & No & 18.6    \\
       level of      &   Yes    &  Yes  & 18.3  \\
       performance?     &       &    &    \\
 \hline
         How insecure,       & No  & No  & 10.0  \\
         discouraged,   &  No     &  Yes  & 11.2  \\
       irritated, stressed,   & Yes      & No & 20.9    \\
       and annoyed     &   Yes    &  Yes  & 37.9  \\
       were you?    &       &     &    \\
       \hline
    \end{tabular}
\end{table}

\subsection{Qualitative Data}
The post-task survey question, \emph{``Could you please type any further comments about the task''}), elicited a range of comments. The following codes were defined, using a grounded theory approach, by the first author and applied to their analysis: 
\vspace{2mm}

\begin{itemize}
    \item irritation/ distraction/ not distracting, 
    \item avoidance,  
    \item enjoyment,  
    \item success/ feedback,  
    \item easy/difficult,  
    \item confusion. 
\end{itemize}
The third author also independently applied the codes, leading to a Cohen's kappa inter-coder agreement of 0.64. Combining the distraction code with the avoidance code led to an agreement of 0.84. The main difference in coding was that two items were coded as being about feedback by one author and difficulty and confusion by the other.

Table~\ref{tab:qualitativeSummary} presents a summary of the qualitative analysis. 
We report below on the results of the analysis as structured by the defined codes.

\paragraph{Distraction} Eight participants (four autistic and four non-autistic participants) commented on the moving logo being distracting, for example, \emph{``The image was too fast rotating; it significantly distracted me.''} One non-autistic participant commented on the layout colours being irritating in the unanimated condition. Two non-autistic participants commented on the unanimated image not distracting them.

\paragraph{Avoidance} Only one comment was associated with this code:  \emph{``I tried to complete this task quickly because of annoying rotating image.''}. The comment was provided by an autistic participant.

\paragraph{Enjoyment} Only one comment was associated with this code: \emph{``enjoyed it''}.  The comment was provided by a non-autistic participant.

\paragraph{Success and Feedback} Two autistic users in the unanimated case and one in the animated case commented on it being \emph{``Hard to gauge success when no search results came up.''} 
Non-autistic users did not comment on this. One in the unanimated case stated \emph{``all was good''}.

\paragraph{Difficulty} Two non-autistic and one autistic participant found the unanimated task easy; similarly for the animated one.

\paragraph{Confusion} One autistic participant stated: \emph{``Easy if I understood it correctly  which I never know if I have or not''}.\\~

\begin{table}[!htb]\centering
    \caption{Summary of the qualitative analysis}
    \label{tab:qualitativeSummary}
    \footnotesize
{\rowcolors{3}{lightgray!30}{white}    
    \begin{tabular}{lcc}
         \hline
        Code &  \multicolumn{2}{c}{No. of comments with each code}  
        \\
         ~& Autistic & Non-Autistic \\ 
         \hline  
         irritation & & 1 (unanimated)
         \\ 
 %        \hline 
         distraction & 4 & 4 \\
%         \\ 
  %       \hline 
         not distracting & & 2 (unanimated)
         \\ 
 %        \hline  
    avoidance & 1 &
    \\ 
 %        \hline  
    enjoyment & & 1
    \\ 
 %        \hline  
      feedback & 3 & 1  
    \\  
 %        \hline  
     difficulty & 1 (easy) & 2 (easy)  
    \\ 
 %        \hline  
     confusion  & 1 &
     \\
       \hline
    \end{tabular}
    }
\end{table}

%========================================

\section{\uppercase{Discussion}}
\label{sec:discussion}

The case of search under the condition of irrelevant animation has led to mixed results. Queries tended to be formulated slightly faster and were more likely to be shorter. Paradoxically, they were also more likely to be pasted, which led to longer queries on average.
The least successful query was typed under an unanimated condition by an autistic participant. However, there was very little difference in query success for this search task across conditions. This was, of course, an artificial situation and the search task description was provided in a manner that allowed copying and pasting. Typed queries were shorter but not slower overall. It would be interesting in a follow-up to find out the reasoning behind participants' choice of typing or pasting their query. It might be that pasting was believed to be faster or possibly easier. It is possible that those who typed their queries did not think of an alternative method, or believed that it wasn't permitted for the study. Participants' decisions may have also depended on their typing skill.

In terms of how users felt during the task, autistic users exposed to animation found it much more mentally demanding than all other cases. Similarly, they were far more irritated by animation than others. Non-autistic users in the non-animated case found the task the least arduous across all cases.

The question on success was answered differently by the two cohorts, but this appears to be less related to a difference in perceived success than a difference in interpretation. The comments from autistic participants revealed a possible reason for this lower success perception. Three autistic participants stated that the lack of feedback after entering the query meant they could not gauge their success.

%====================================
\section{\uppercase{Limitations}}
\label{sec:limit}

The purpose of the full study was to determine the effect of on-screen animations on users, with particular focus on how this impacted autistic users, who are known to have different responses to sensory input than the general population. What is reported here is on the search part of the overall study, and is therefore not a comprehensive look at animation and search. The participants did not receive feedback in the form of search results and there was only a single search task. The number of participants was small with considerable variation in their responses, meaning a purely statistical analysis would be unjustified and extrapolating to all users would be risky. 

It is possible that the differences between the autistic and non-autistic cohorts would have been greater if all participants had a similar language background. An artifact of recruitment was that the non-autistic cohort were students, the majority of whom did not have English as their first language, whereas this did not appear to be the case for the autistic cohort. Future studies are expected to account for this variable.
It should also be noted that the autistic participants were without cognitive impairment (Autism Spectrum Condition level 1) and that those at level 2 may be differently impacted.

%===================================

\section{\uppercase{Conclusions}}
\label{sec:concl}
In this paper we presented an analysis of the impact of animated user interface elements on Web-based search activities. The analysis
examined how two cohorts, autistic and non-autistic users, create  search queries for a specific search task in the presence or absence of on-screen animations. Our aim was to answer the following research question:

\emph{\textbf{RQ:} How does the effect of irrelevant animations within user interfaces vary between autistic and non-autistic users,  while conducting Web search tasks?} 

Based on the results of our experiments, we can conclude that  
there was little practical difference in the time taken or the success of the query, but notable differences in strategy of query formulation, with more people pasting their queries in the presence of animation. In addition, autistic users found the task more mentally demanding and irritating than non-autistic users when animation was present. 
Thus, even for short tasks such as formulating a search query, it would be beneficial to avoid having any irrelevant animated elements in Web interfaces. 

A possible future work direction is to replicate the study with more queries and different types of animation, e.g., flashing, because each type is likely to have different effects on users, particularly autistic users. Some types of animation, such as flashing, are probably even more distracting than those used in our study, and identification of these potential issues might be useful for making software more inclusive and accessible.  It would be also interesting to analyse whether the cultural or gender  aspects might have any influence on the results of the study, so it might be useful to expand this study with the analysis replicating the works of  \cite{alharthi2018gender,alharthi2019social} and   \cite{alsanoosy2020exploratory,alsanoosy2020identification} in the settings of our study.

%===================================
%\vfill
\section*{\uppercase{Acknowledgements}}

The web application was adapted from a student programming project.

We thank Gabrielle Hall
for her input into the experimental design and insights into autistic research. 
We also thank all participants for the time they invested in our study.

%\newpage
%===================================
\bibliographystyle{apalike}
%{\small
%\bibliography{main} 
%}

\end{document}